\documentclass[a4paper]{jpconf}
\usepackage{graphicx}

\newcommand{\be}{\begin{equation}}
\newcommand{\ee}{\end{equation}}

\newcommand{\ket}{\rangle}
\newcommand{\bea}{\begin{eqnarray}}
\newcommand{\eea}{\end{eqnarray}}

\usepackage{physics}
\usepackage{quantikz}
\usepackage{url}

\begin{document}
\title{Volatility time series modeling by single-qubit quantum circuit learning
}

\author{Tetsuya Takaishi}

\address{Hiroshima University of Economics, Hiroshima 731-0192, JAPAN}

\ead{tt-taka@hue.ac.jp}

\begin{abstract}
We employ single-qubit quantum circuit learning (QCL) to model the dynamics of volatility time series. To assess its effectiveness, we generate synthetic data using the Rational GARCH model, which is specifically designed to capture volatility asymmetry.  
Our results show that QCL-based volatility predictions preserve the negative return–volatility correlation, a hallmark of asymmetric volatility dynamics.  
Moreover, analysis of the Hurst exponent and multifractal characteristics indicates that the predicted series, like the original synthetic data, exhibits anti-persistent behavior and retains its multifractal structure.  
\end{abstract}

\section{Introduction}

In finance, volatility is a key indicator used to measure the magnitude of fluctuations in the time series of financial assets, and is commonly employed as a risk measure.  
For practitioners, predicting future volatility is an essential task for managing the risk of held financial assets and mitigating potential losses.  
A common approach to forecasting volatility involves introducing time series models that replicate the statistical properties observed in financial data.  
Therefore, constructing appropriate volatility models requires a prior understanding of the empirical characteristics inherent in financial time series.  

According to various statistical studies, financial time series exhibit universal properties across different asset classes.  
These properties are collectively referred to as "stylized facts"\cite{Cont2001QF}.  
One notable stylized fact is volatility clustering, which refers to the tendency for periods of high and low volatility to occur in succession.  
A widely used model that captures this property is the generalized autoregressive conditional heteroscedasticity (GARCH) model\cite{Bollerslev1986JOE}, which has been extensively applied in empirical research.  

A limitation of the GARCH model is its inability to capture volatility asymmetry.  
Empirical studies have shown that volatility—particularly in equity markets—tends to increase more following negative returns than positive ones, a phenomenon known as the leverage effect\cite{Black1976}.

This effect can also be observed through the return–volatility correlation, where returns are negatively correlated with future volatility\cite{bouchaud2001leverage,roman2008skewness,takaishi2020power}\footnote{Empirical studies have reported the presence of an anti-leverage effect in the Chinese market\cite{qiu2006return,shen2009return}.}.  
To accommodate volatility asymmetry, several extensions of the GARCH model have been proposed, including the Exponential GARCH\cite{Nelson1991Econ}, Quadratic GARCH\cite{Sentana1995RES}, GJR GARCH\cite{Glosten1993JOF}, Asymmetric Power GARCH\cite{ding1993long} and Rational GARCH (RGARCH)\cite{takaishi2017rational,takaishi2018volatility} models.  

Recently, Gatheral et al.\cite{gatheral2018volatility} found that the Hurst exponent of volatility increment time series is less than 0.5, indicating anti-persistent behavior and a rough volatility path.  
Volatility exhibiting these properties is referred to as "rough volatility."  
Focusing on this feature, they proposed a volatility model based on fractional Brownian motion.  
Subsequent empirical studies have confirmed that volatility time series exhibit characteristics of rough volatility\cite{bennedsen2022decoupling,livieri2018rough,takaishi2020rough,floc2022roughness,brandi2022multiscaling,takaishi2025multifractality}.  

Although numerous volatility models have been proposed, the estimated volatility values vary across models, necessitating careful consideration in model selection.  
One approach to circumvent this issue is to use realized volatility (RV)\cite{andersen1998answering,mcaleer2008realized}, a model-free measure calculated from high-frequency intraday data.  
While RV does not require model specification for its computation, forecasting volatility inevitably involves choosing a model, which reintroduces ambiguity.  
Among the models for RV, the heterogeneous autoregressive (HAR) model\cite{corsi2009simple} and realized stochastic volatility (RSV) models\cite{takahashi2009estimating,takaishi2014RSV,takaishi2015GPU, takaishi2015application,takaishi2018bias}, realized GARCH model\cite{hansen2012realized} are well known.  

In this study, we model volatility time series using quantum circuit learning (QCL)\cite{mitarai2018quantum}, a classical–quantum hybrid algorithm capable of approximating nonlinear functions.  
In QCL, a parameterized quantum circuit (PQC) is introduced, and its parameters are optimized to minimize a loss function defined by the difference between the teacher data and the PQC outputs.  
The advantage of QCL in volatility modeling lies in its flexibility: it does not require any prior assumptions about the properties of volatility.  
Instead, QCL automatically generates a nonlinear function that approximates the volatility structure by tuning the PQC parameters.  

We employ synthetic financial data generated by the RGARCH model, which produces asymmetric volatility time series.  
Using these data, we investigate the feasibility of volatility modeling via QCL, with particular emphasis on capturing volatility asymmetry and multifractal characteristics.

\section{Quantum Circuit Learning for Volatility Modeling}

In the QCL framework,  
we aim to learn a volatility function $v_i = v(\boldsymbol{x}_i, \boldsymbol{\theta})$  
that approximates the target volatility data (i.e., teacher data) $\sigma_i^2$  
by tuning the parameter vector $\boldsymbol{\theta}$.  
The input data is defined as $\boldsymbol{x}_i = (r_i, \sigma_i^2)$,  
where $r_i$ denotes the return and $\sigma_i^2$ the corresponding volatility.

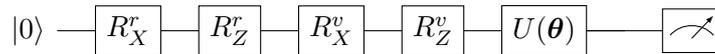
\begin{figure}[h]
\centering
\begin{quantikz}[thin lines]
  \lstick{{$\ket{0}$}} & \gate{R_Y^r} & \gate{R_Z^r} & \gate{R_Y^v} & \gate{R_Z^v} & \gate{U(\boldsymbol{\theta})} & \qw & \meter{} \\
\end{quantikz}
\caption{Single-qubit quantum circuit used for volatility modeling}
\label{fig:volatility_circuit}
\end{figure}

Figure~\ref{fig:volatility_circuit} illustrates the single-qubit PQC employed in this study.

The unitary gate $U(\boldsymbol{\theta})$ is a two-dimensional operator defined as
\[
U(\boldsymbol{\theta}) = 
\begin{pmatrix}
   \cos (\theta/2) & -e^{i\lambda} \sin (\theta/2)  \\
   e^{i\phi}\sin(\theta/2) & e^{i(\lambda + \phi)}\cos(\theta/2)
\end{pmatrix},
\]
where $\boldsymbol{\theta} = (\theta, \lambda, \phi)$ represents the trainable parameters.

Angle encoding is applied to embed the input data into quantum states using rotational gates, following the method proposed by Mitarai et al.~\cite{mitarai2018quantum}:
\begin{align}
R_Y^r &= R_Y(\arcsin(r_i)) \\
R_Z^r &= R_Z(\arccos(r_i^2)) \\
R_Y^v &= R_Y(\arcsin(2\sigma_i^2 - 1)) \\
R_Z^v &= R_Z(\arccos(\sigma_i^4))
\end{align}

The output is obtained via a Z-basis measurement, and the probability $P_0$ of measuring the $\ket{0}$ state is used to represent the predicted volatility at time $i+1$, i.e., $v_{i+1} = P_0$.
In the actual simulation, we utilize Qiskit\footnote{IBM Qiskit: \url{https://www.ibm.com/quantum/qiskit}} to compute the probability $P_0$ directly from the state vector.

The optimal parameter set $\boldsymbol{\theta}$ is determined by minimizing the following loss function:
\[
L = \sum_{i=1}^{N} (v_i - \sigma_i^2)^2,
\]
where $N$ is the number of data points.

\section{Rational GARCH Model}

The RGARCH model\cite{takaishi2017rational} is designed to capture volatility asymmetry and is defined as
\be 
\sigma_t^2 = \frac{ \omega + \alpha r_{t-1}^2 + \beta \sigma_{t-1}^2 }{1 + \gamma r_{t-1}},
\label{eq:rgarch}
\ee 
\be 
r_t = \sigma_t \epsilon_t,
\ee 
where $r_t$ and $\sigma_t^2$ denote the return and volatility at time $t$, respectively.  
The parameters $\alpha$, $\beta$, $\omega$, and $\gamma$ govern the model dynamics, with $\gamma$ controlling the degree of volatility asymmetry.  
When $\gamma = 0$, the RGARCH model reduces to the standard GARCH model, which assumes symmetric volatility.  
Here, $\epsilon_t$ represents a Gaussian random variable with mean zero and unit variance.

The denominator in Eq.~(\ref{eq:rgarch}) may become negative when $\gamma$ is large and/or the magnitude of returns is high, which violates the non-negativity condition of volatility.  
To address this issue, an improved volatility equation has been proposed\cite{takaishi2018volatility}, given by
\be 
\sigma_t^2 = \frac{ \omega + \alpha r_{t-1}^2 + \beta \sigma_{t-1}^2 }{\exp(\gamma r_{t-1})},
\label{eq:exp}
\ee 
which converges to the GARCH model when $\gamma = 0$.

To examine the leverage effect in the RGARCH model, we compute the return–volatility cross-correlation\cite{takaishi2020power} with time lag $j$, defined as
\be 
C_d(j) = \frac{E[(r_t - \mu_r)(\sigma_{t+j}^d - \mu_{\sigma^d})]}{S_r S_{\sigma^d}},
\label{eq:CC}
\ee 
where $\mu_r$ and $\mu_{\sigma^d}$ denote the mean values of $r_t$ and $\sigma_t^d$, respectively, and $S_r$ and $S_{\sigma^d}$ are their corresponding standard deviations.  
The operator $E[\cdot]$ denotes the expectation value.  
Eq.~(\ref{eq:CC}), inspired by the study of the Taylor effect\cite{takaishi2018taylor}, is a generalized form of the correlation function used in previous studies\cite{bouchaud2001leverage,roman2008skewness}.

\begin{figure}[h]
\centering
\includegraphics[width=8.5cm]{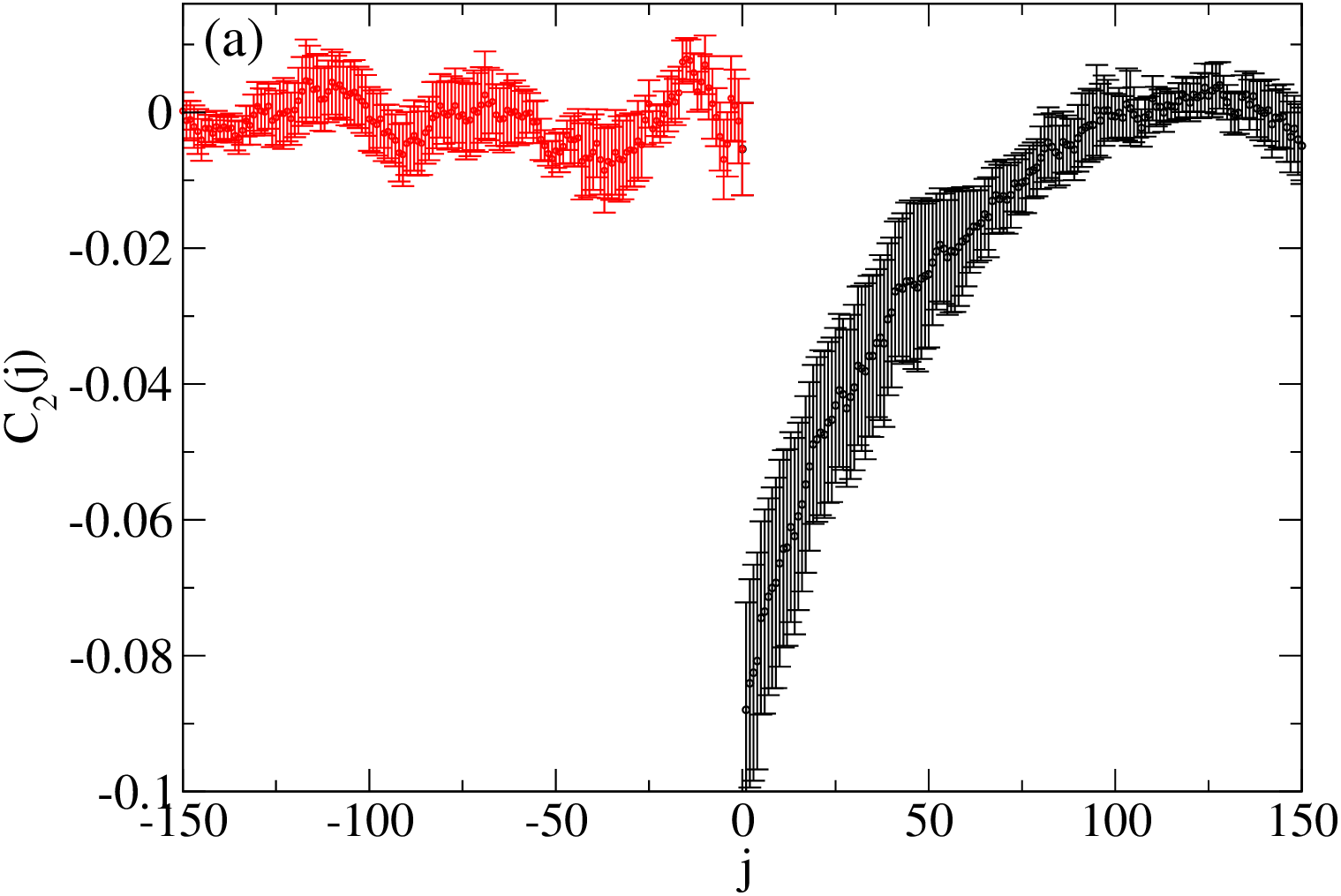} \\
\vspace{10mm}
\includegraphics[width=8.5cm]{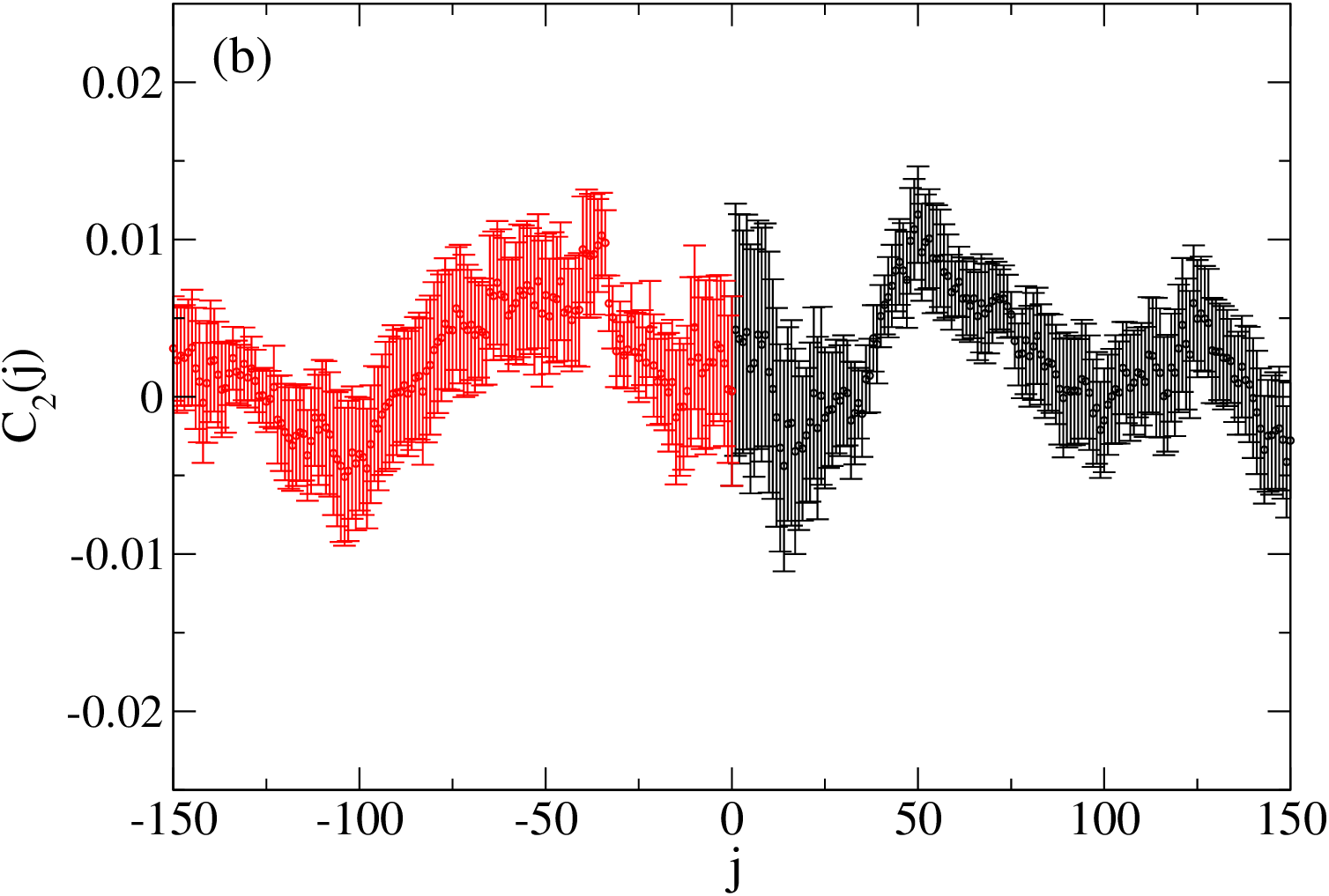}
\caption{\label{fig:rgarch_corr}
(a) $C_2(j)$ computed from RGARCH data with asymmetric volatility $(\alpha = 0.11, \beta = 0.85, \omega = 0.005, \gamma = 0.1)$.  
(b) $C_2(j)$ computed from symmetric volatility data $(\alpha = 0.11, \beta = 0.85, \omega = 0.005, \gamma = 0.0)$.}
\end{figure}

We generate return and volatility time series using the volatility function in Eq.~(\ref{eq:exp}) with parameters $(\alpha = 0.11, \beta = 0.85, \omega = 0.005, \gamma = 0.1)$.  
Figure~\ref{fig:rgarch_corr} (a) shows $C_2(j)$ for $d = 2$, calculated from 100{,}000 simulated data points.  
For positive values of $j$, $C_2(j)$ exhibits negative correlation, indicating that returns are negatively correlated with future volatility.  
In contrast, for negative values of $j$, no significant return–volatility correlation is observed.  
Similar patterns—namely, negative correlations for positive $j$—are also found for other values of $d$.

To compare with the symmetric volatility case ($\gamma = 0.0$),  
Figure~\ref{fig:rgarch_corr} (b) presents $C_2(j)$ computed from 1{,}000{,}000 data points generated using the same parameters except for $\gamma = 0.0$.  
As shown in the figure, the return–volatility correlation disappears under symmetric volatility conditions.

\section{Volatility Modeling via Quantum Circuit Learning}

\begin{figure}[h]
\centering
\includegraphics[width=11cm]{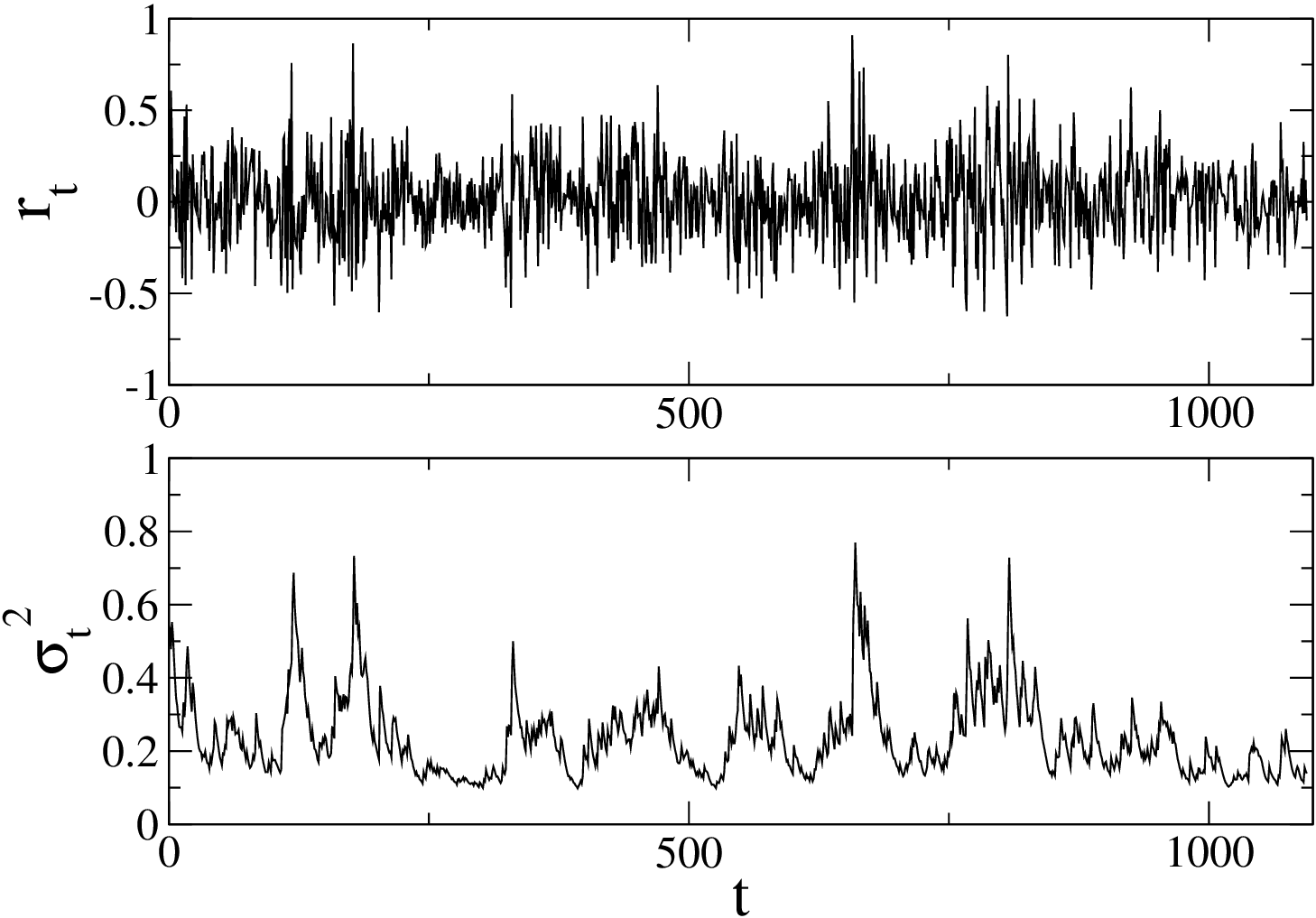}
\caption{Return $r_t$ and volatility $\sigma_t^2$ used as input data.}
\label{fig:input_data}
\end{figure}

We first generate 1,095 data points (corresponding to three years of daily data) using the RGARCH model with parameters $(\alpha = 0.11, \beta = 0.85, \omega = 0.005, \gamma = 0.1)$.  
Figure~\ref{fig:input_data} shows the returns $r_t$ and volatilities $\sigma_t^2$ used as input data.  
The data are scaled to ensure that their absolute values do not exceed 1, as they are employed as angular components in the angle encoding.

\begin{figure}[]
\vspace{10mm}
\centering
\includegraphics[width=10cm]{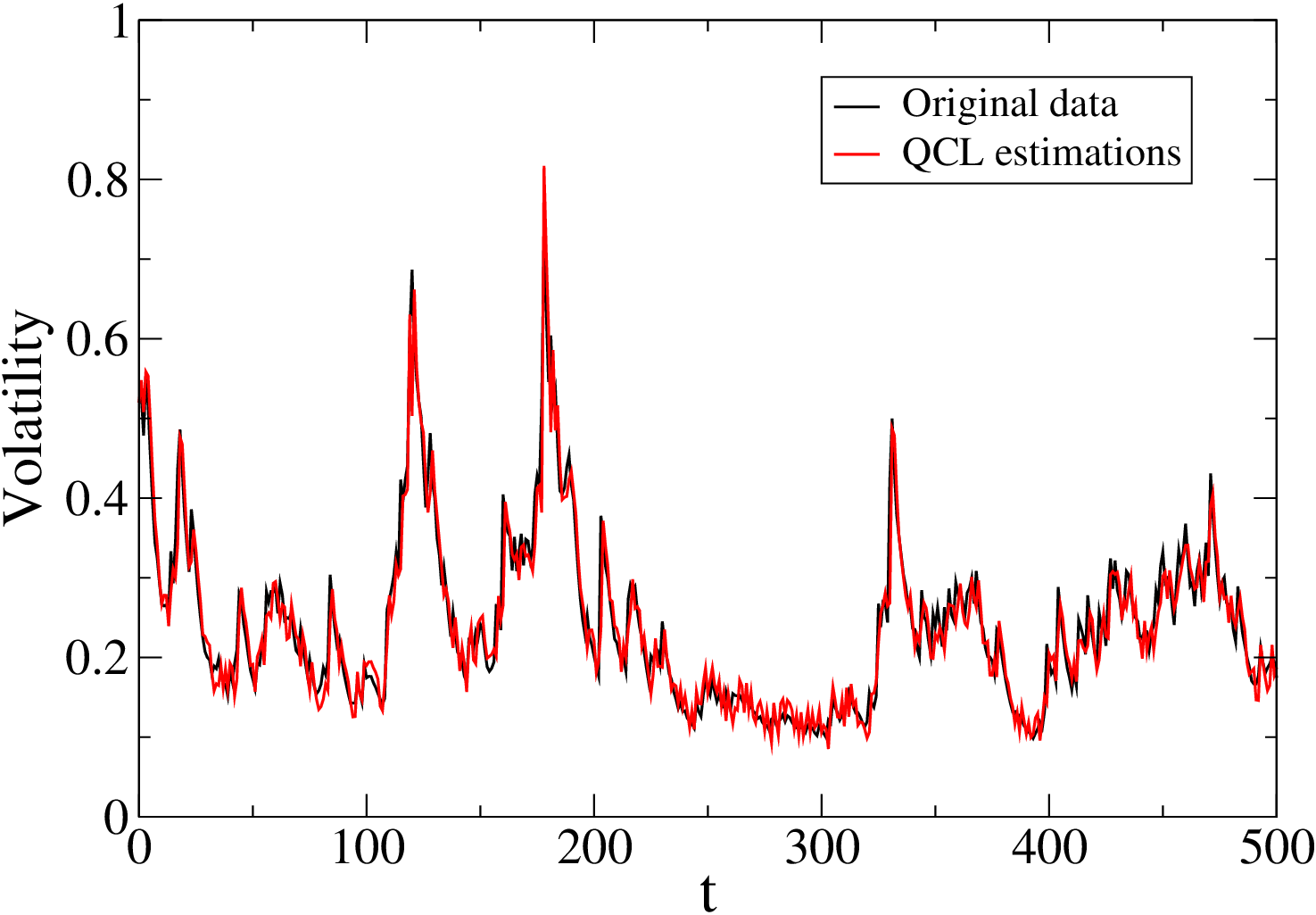}
\caption{Comparison between the original volatility and the volatility estimates obtained via QCL. The first 500 data points are displayed.}
\label{fig:qcl_estimation}
\end{figure}

\begin{figure}[]
\centering
\includegraphics[width=11cm]{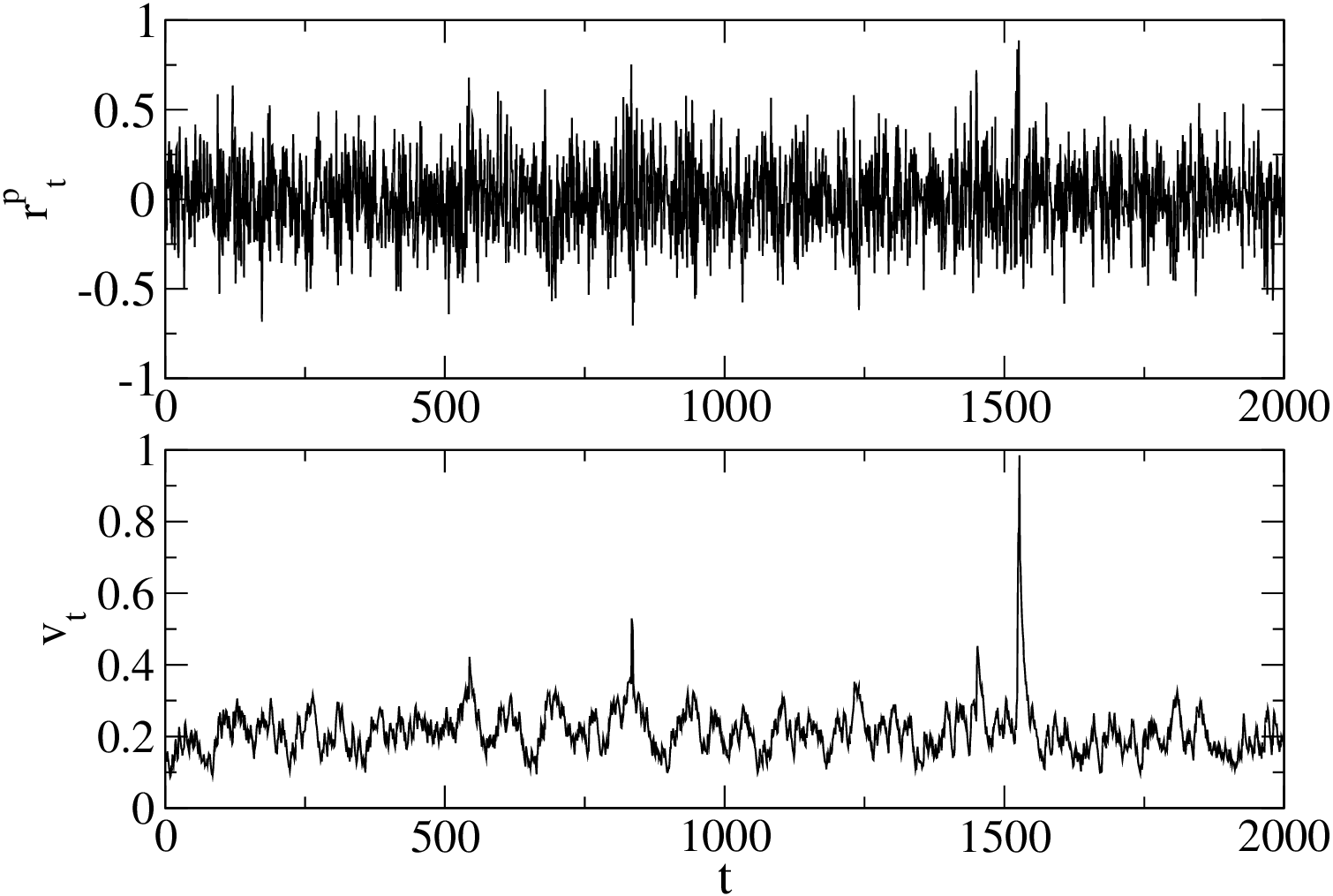}
\caption{Predicted return $r^p_t$ and volatility $v_t$ generated using QCL.}
\label{fig:qcl_prediction}
\end{figure}

\begin{figure}[]
\vspace{10mm}
\centering
\includegraphics[width=9.5cm]{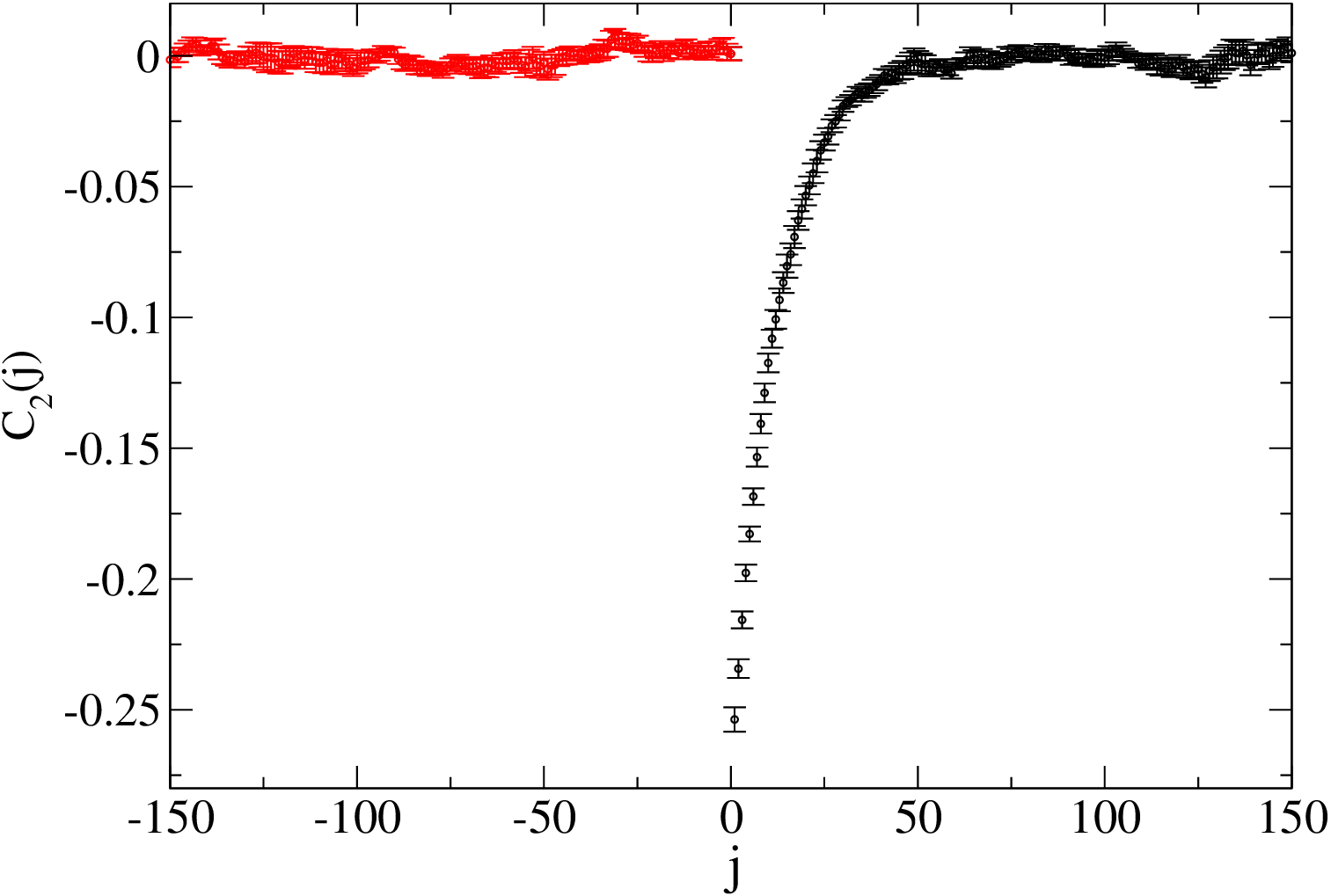}
\caption{$C_2(j)$ computed from QCL-generated data.}
\label{fig:c2_function}
\end{figure}

Using the rescaled data, we estimate volatilities via QCL with the single-qubit quantum circuit shown in Figure~\ref{fig:volatility_circuit}, optimizing the parameters to minimize the loss function.  
The QCL simulation is performed using Qiskit, and the optimization is carried out using the COBYLA algorithm.  
Figure~\ref{fig:qcl_estimation} compares the QCL-based volatility estimates with the original volatility data.  
The results indicate that the QCL estimates closely follow the fluctuations observed in the original data.

\begin{figure}[]
\centering
\includegraphics[width=8.5cm]{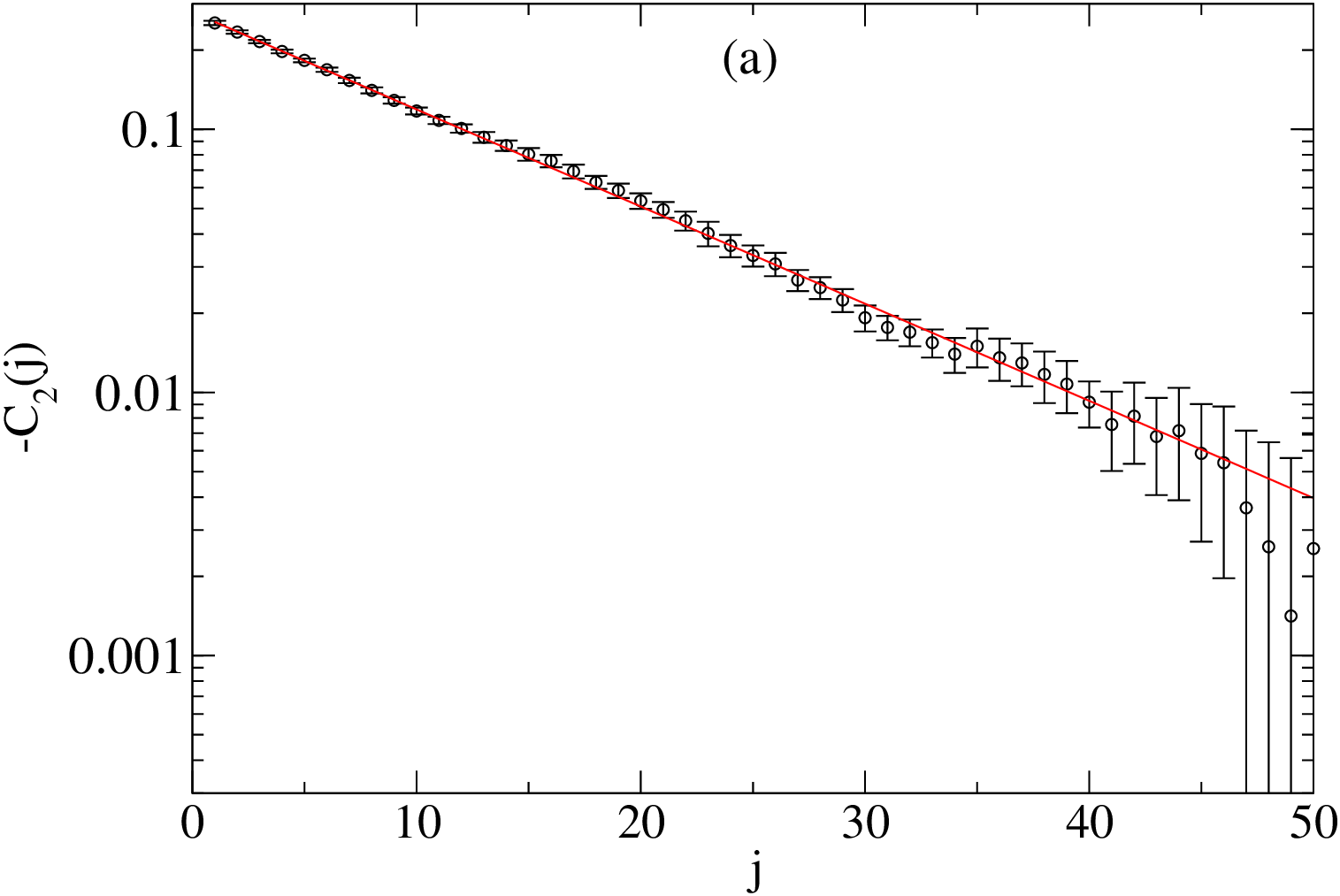} \\
\vspace{10mm}
\includegraphics[width=8.5cm]{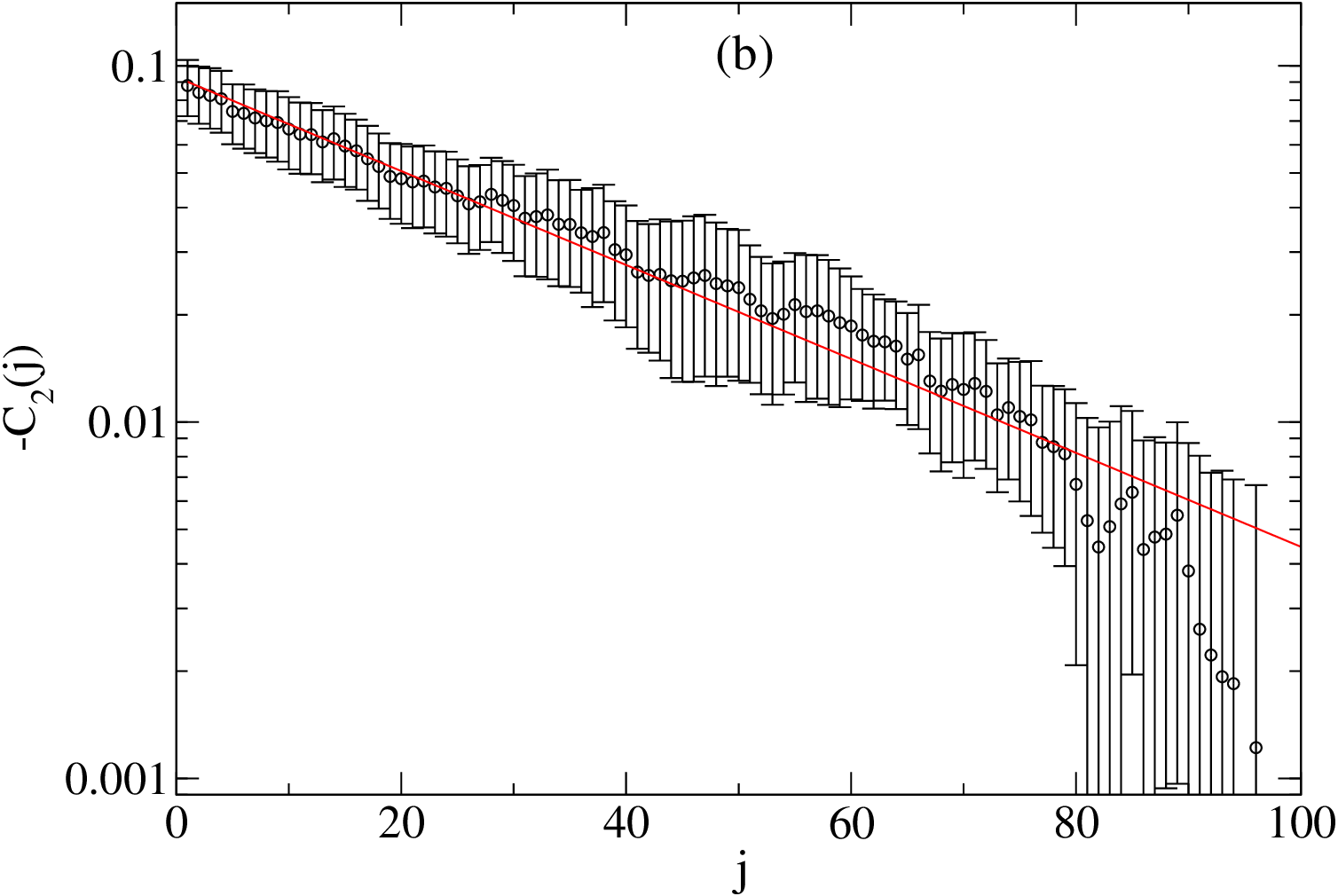}
\caption{(a) $-C_2(j)$ derived from QCL predictions, with an exponential fit (red line) to $\sim \exp(-j/\tau)$, yielding $\tau \approx 12$.  
(b) $-C_2(j)$ derived from the original data. The exponential fit yields $\tau \approx 33$.}
\label{fig:exp_decay}
\end{figure}

Next, using the optimized parameters from QCL, we generate 100{,}000 data points for the predicted return $r^p_t$ and volatility $v_t$, where $r^p_t = \sqrt{v_t} \epsilon_t$.  
Figure~\ref{fig:qcl_prediction} plots the first 2{,}000 predictions of $r^p_t$ and $v_t$, showing similar variations to the original data in Figure~\ref{fig:input_data}.

Figure~\ref{fig:c2_function} shows $C_2(j)$ calculated from the 100{,}000 QCL-generated data points.  
As with the original data, $C_2(j)$ exhibits negative values for positive $j$ and no significant correlation for negative $j$.

In Figure~\ref{fig:exp_decay}, $-C_2(j)$ is plotted on a semi-logarithmic scale to examine the correlation decay.  
The figure demonstrates that $-C_2(j)$ exhibits exponential decay in both the original and QCL-generated datasets.  
This behavior is consistent with previous findings\cite{bouchaud2001leverage}, whereas high-frequency Bitcoin data exhibit power-law decay\cite{takaishi2020power}, which deviates from the RGARCH results.

\begin{figure}[]
\centering
\includegraphics[width=10cm]{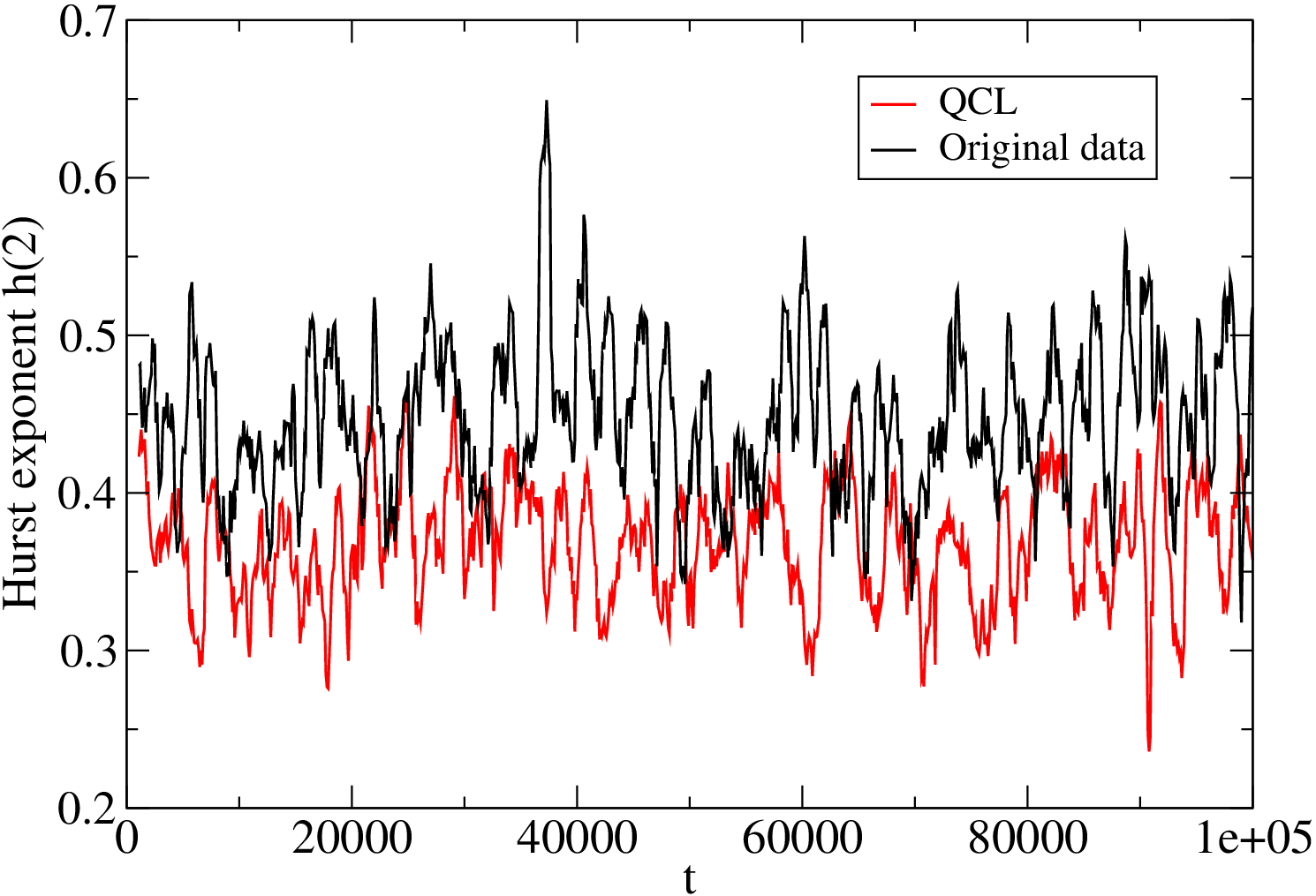}
\caption{Hurst exponent $h(2)$ computed from QCL and original data.}
\label{fig:hurst}
\end{figure}

\begin{figure}[]
\centering
\includegraphics[width=8.5cm]{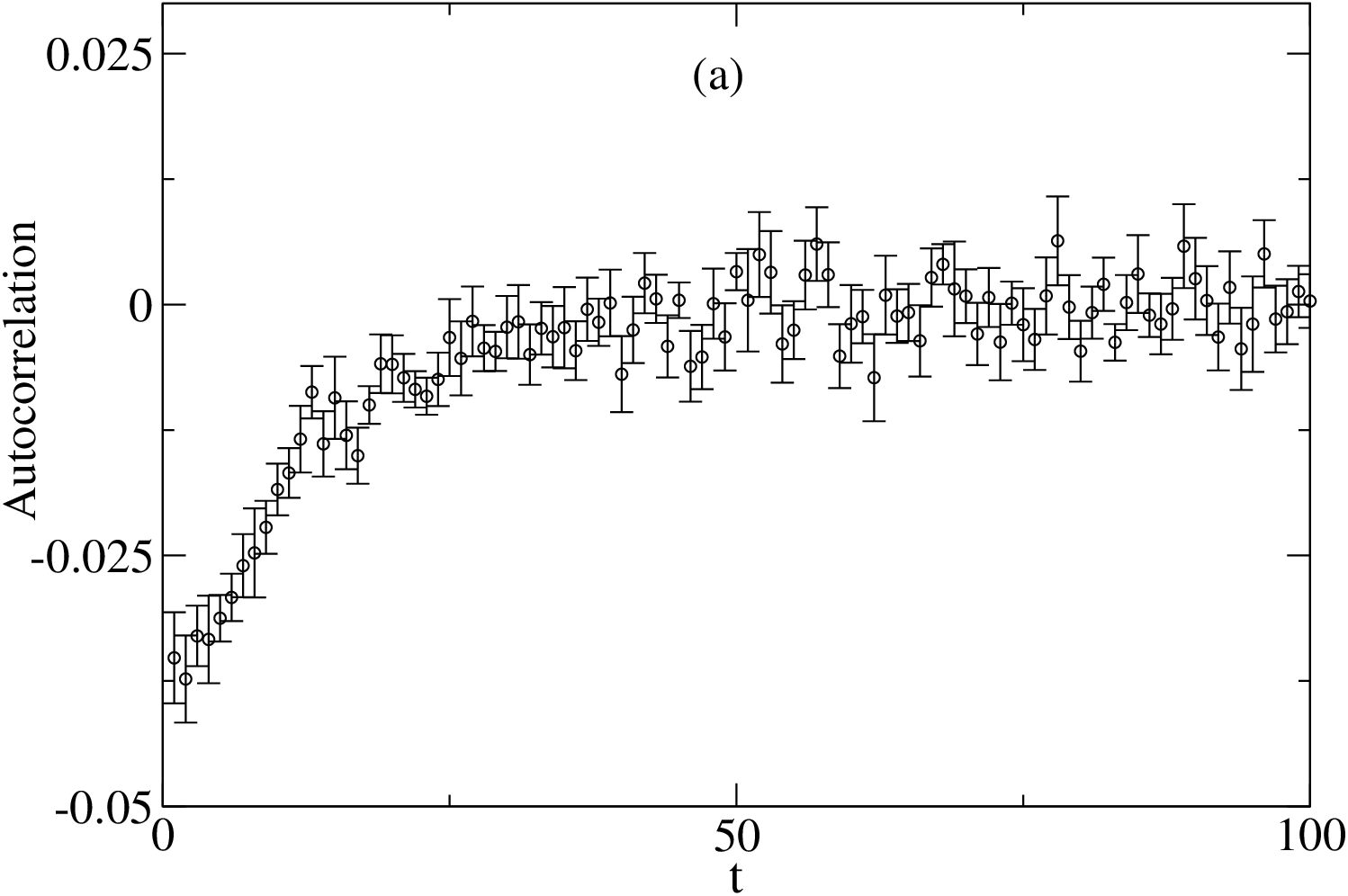} \\
\vspace{10mm}
\includegraphics[width=8.5cm]{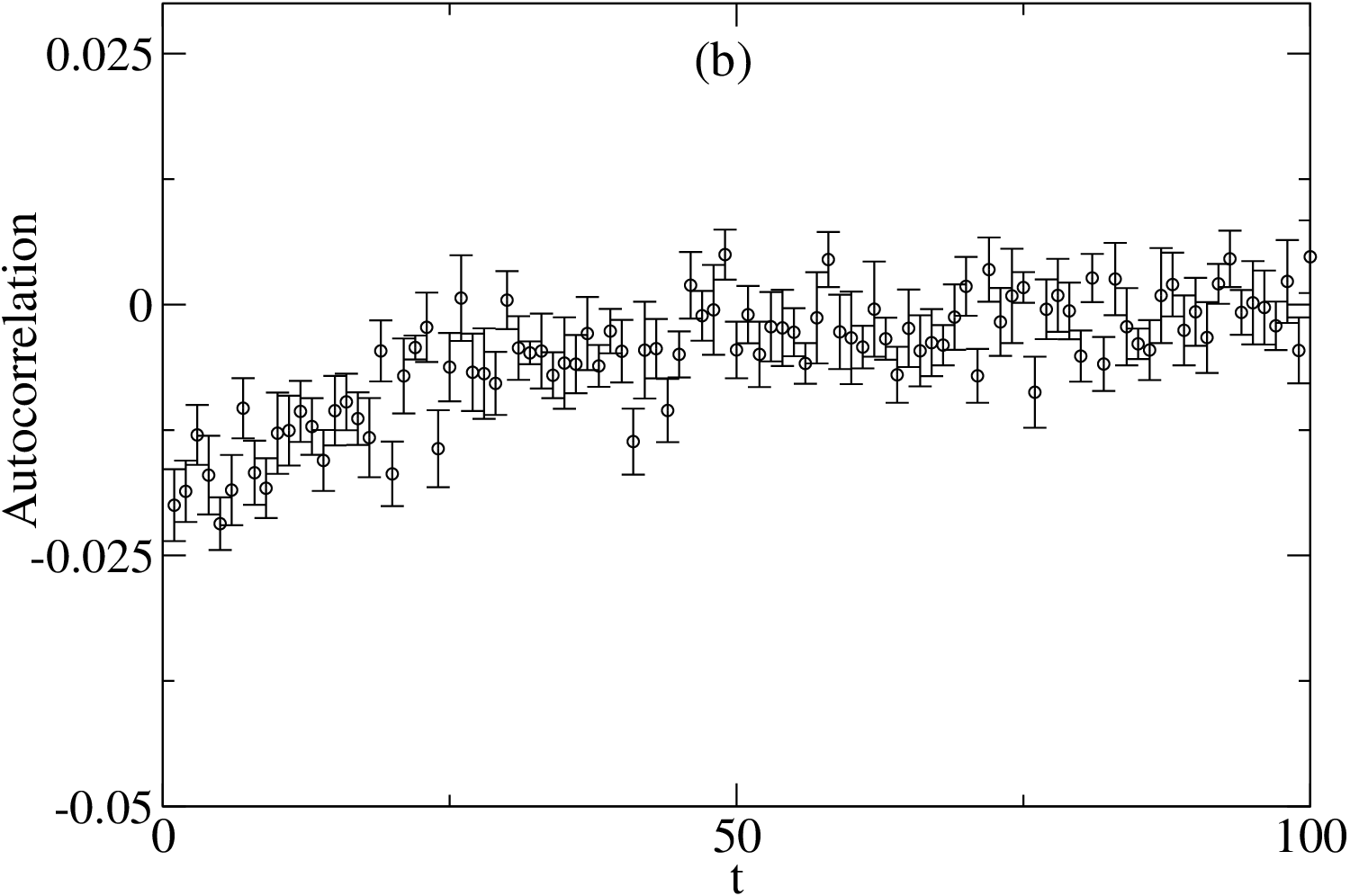}
\caption{(a) Autocorrelation function from QCL predictions.  
(b) Autocorrelation function from original data.}
\label{fig:autocorr}
\end{figure}

\begin{figure}[]
\centering
\includegraphics[width=8.5cm]{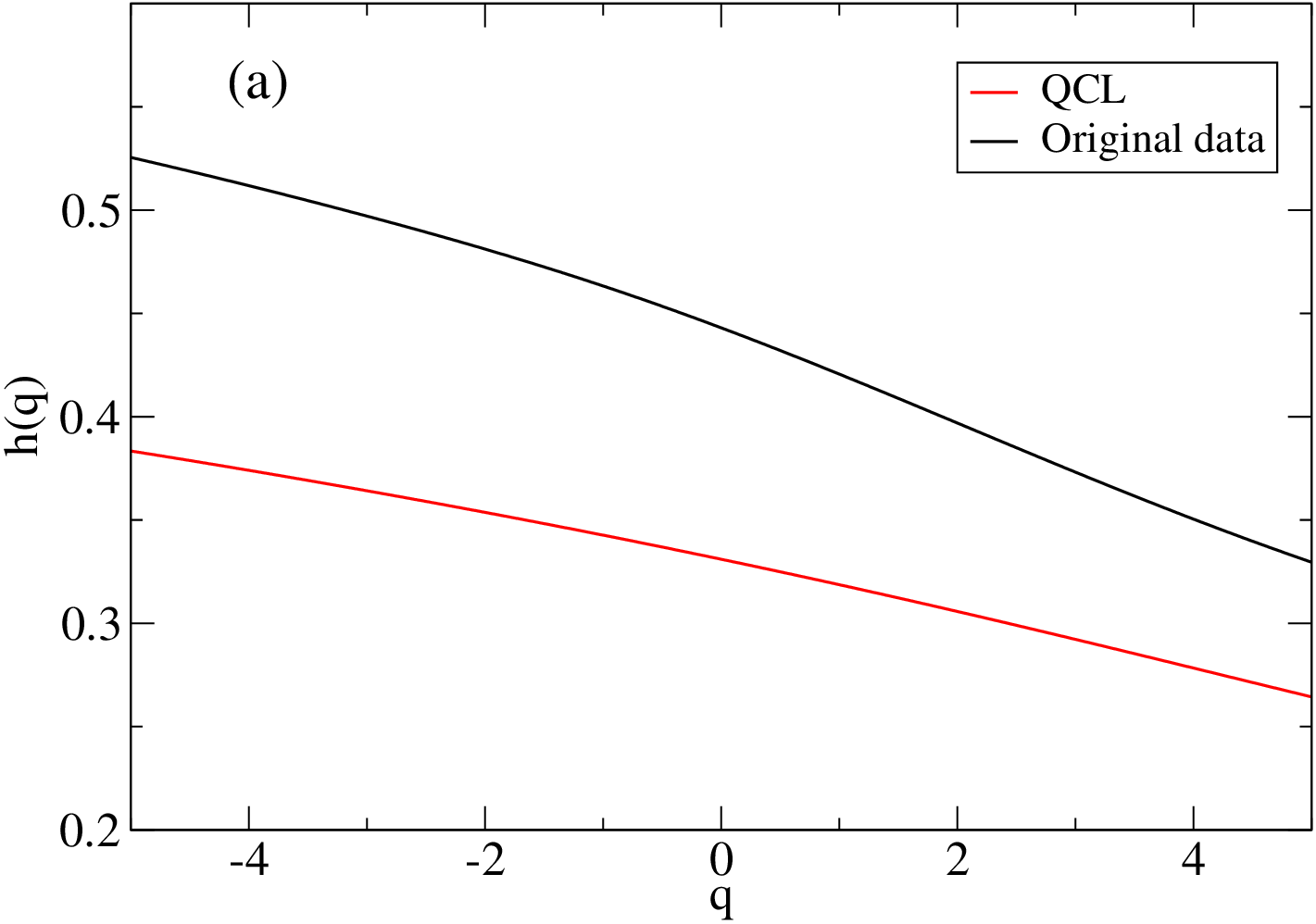}\\
\vspace{10mm}
\includegraphics[width=8.5cm]{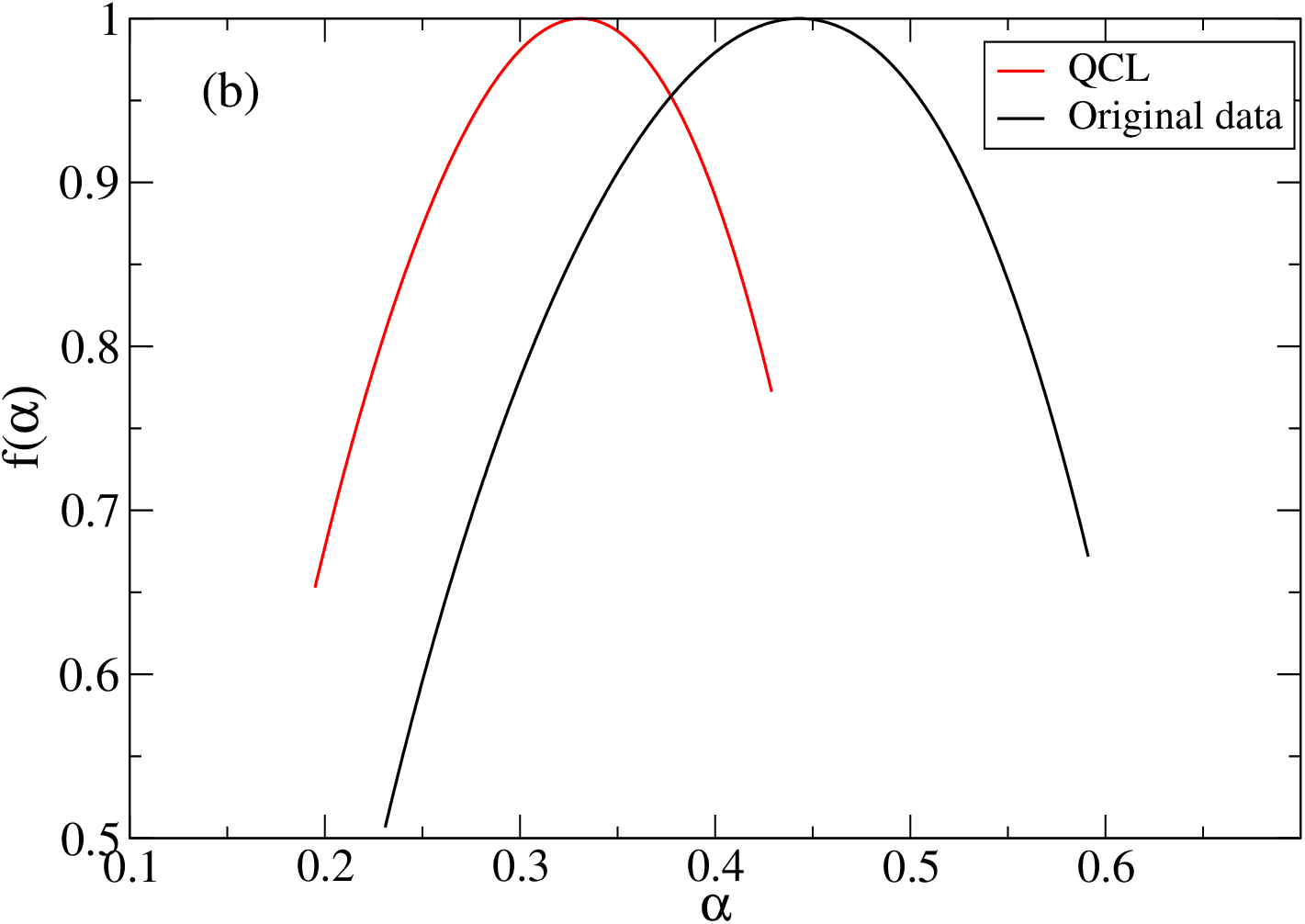}
\caption{(a) Generalized Hurst exponent $h(q)$.  
(b) Singularity spectrum $f(\alpha)$.}
\label{fig:multifractal}
\end{figure}

Empirically, volatility increment time series are known to exhibit anti-persistent behavior, with a Hurst exponent less than 0.5.  
The volatility increment $dV_t$ is defined as $dV_t = \log v_t - \log v_{t-1}$.  
We compute the Hurst exponent for both the original and QCL-generated data using multifractal detrended fluctuation analysis (MDFA)\cite{kantelhardt2002multifractal} over segments of length 1,095.  
To examine temporal variation, we apply a rolling window approach with a shift of 100 data points.  
Figure~\ref{fig:hurst} shows the Hurst exponent $h(2)$ for both datasets, with average values below 0.5, confirming anti-persistent behavior.

Figure~\ref{fig:autocorr} presents the autocorrelation function of the volatility increment time series, which shows negative values near the origin.  
This negative autocorrelation further supports the presence of anti-persistent behavior.

A time series is considered multifractal when the generalized Hurst exponent $h(q)$ varies with respect to $q$.  
Figure~\ref{fig:multifractal} (a) shows $h(q)$ computed over a segment of 1,095 observations.  
For both QCL-generated and original data, $h(q)$ is non-constant, indicating multifractal behavior.  
Figure~\ref{fig:multifractal} (b) presents the singularity spectrum $f(\alpha)$\cite{kantelhardt2002multifractal}, which spans a broad range of $\alpha$ values in both datasets, further confirming multifractality.

\section{Conclusion}

In this study, we investigated the feasibility of modeling volatility time series using quantum circuit learning (QCL) with a single-qubit parameterized quantum circuit.  
By employing angle encoding and optimizing circuit parameters to minimize the loss function, we demonstrated that QCL can effectively approximate the nonlinear structure of volatility generated by the RGARCH model, which captures asymmetric volatility behavior.

Our results show that QCL-based volatility estimates closely follow the original data and successfully reproduce key statistical properties, including the leverage effect, exponential decay of return-volatility correlation, and anti-persistent behavior as indicated by the Hurst exponent.  
Furthermore, both the original and QCL-generated time series exhibit multifractal characteristics, as confirmed by the generalized Hurst exponent and singularity spectrum analyses.

These findings suggest that QCL provides a promising framework for volatility modeling without requiring prior assumptions about the underlying dynamics.  
The ability of a single-qubit quantum circuit to capture complex temporal patterns highlights the potential of quantum machine learning in financial time series analysis.  
Future work may extend this approach to multi-qubit architectures, data re-uploading models \cite{perez2020data,perez2021one}, and real-world financial datasets in order to assess the practical utility of the QCL-based volatility model.

\section*{Acknowledgements}
The numerical calculations in this study were carried out using the computer facilities of the Yukawa Institute for Theoretical Physics and the Institute of Statistical Mathematics.  
This work was supported by the Yu-cho Foundation (Grant-in-Aid for Research, 2024).
\section*{References}

\bibliographystyle{plain}
\bibliography{ICMSQ2025QCLv3.bbl}

\end{document}